\documentclass[reprint,superscriptaddress,amsmath,amssymb,aps,notitlepage,floatfix,longbibliography]{revtex4-2}
\usepackage{graphicx}
\usepackage{color}
\usepackage{amsmath}
\usepackage{mathtools}
\usepackage{amssymb}

\usepackage{dcolumn}
\usepackage{upgreek}
\usepackage{gensymb}
\usepackage{float}
\usepackage{bm}
\usepackage{xcolor}
\usepackage{comment}
\usepackage{here}
\usepackage{tabularx}
\usepackage{booktabs}
\usepackage{braket}

\usepackage{siunitx}
\DeclareSIUnit\Oersted{Oe}
\DeclareSIUnit\electronvolt{eV}

\usepackage{xr}
\makeatletter
\newcommand*{\addFileDependency}[1]{
  \typeout{(#1)}
  \@addtofilelist{#1}
  \IfFileExists{#1}{}{\typeout{No file #1.}}
}
\makeatother

\renewcommand{\selectlanguage}[1]{}

\begin{document}

	
\title{Current induced magneto-optical Kerr effect as a probe of Dirac carriers in Bi$_{1-x}$Sb$_x$ alloy}

\author{Ryota Miyazaki}
\affiliation{Department of Physics, The University of Tokyo, Tokyo 113-0033, Japan}

\author{Shunzhen Wang}
\affiliation{Department of Physics, The University of Tokyo, Tokyo 113-0033, Japan}

\author{Guanxiong Qu}
\affiliation{Department of Physics and Astronomy, University of California, Irvine, CA 92697, USA}
\affiliation{RIKEN Center for Emergent Matter Science (CEMS), Wako 351-0198, Japan}

\author{Yukihiro Marui}
\affiliation{Research Institute of Electrical Communication (RIEC), Tohoku University, Sendai 980-8577, Japan}

\author{Yuta Kobayashi}
\affiliation{Department of Physics, The University of Tokyo, Tokyo 113-0033, Japan}

\author{Masashi Kawaguchi}
\affiliation{Department of Physics, The University of Tokyo, Tokyo 113-0033, Japan}


\author{Masamitsu Hayashi}
\affiliation{Department of Physics, The University of Tokyo, Tokyo 113-0033, Japan}
\affiliation{Trans-scale Quantum Science Institute, The University of Tokyo, Tokyo 113-0033, Japan}

\newif\iffigure
\figurefalse
\figuretrue

\date{\today}

\begin{abstract}
We study the current-induced magneto-optical Kerr effect (MOKE) in Bi$_{1-x}$Sb$_x$ semi-metalic alloys. The MOKE signal is found to be the largest in pure Bi ($x=0$), exceeding that of transition metals by nearly four orders of magnitude, and decreases monotonically with increasing Sb concentration. 
We find the MOKE signal scales with the resistivity ($\rho$) as $\rho^{1.7 \pm 0.6}$ and with the mobility ($\mu_\mathrm{c}$) as $\mu_\mathrm{c}^{2.0 \pm 0.2}$.
Model calculations show that such exponent can be accounted for if the Dirac electrons are responsible for the generation of spin current.
This is in contrast to the $\rho^{2}$ and $\mu_\mathrm{c}^{-2}$ scaling of the MOKE signal induced by the free electrons in parabolic band.
The scaling of the MOKE amplitude with the resistivity also partly accounts for the order of magnitude differences of the signal observed between metals, semimetals, and semiconductors.
These results demonstrate that current induced MOKE serves as an effective means to characterize the nature of spin current in materials with diverse electronic structures.
\end{abstract}

\maketitle
The magneto-optical Kerr effect (MOKE) can probe accumulation of spin magnetic moments in semiconductors and metals induced by the spin Hall effect\cite{kato2004science,vanterve2014apl,stamm2017prl,yokouchi2021prap,marui2023prb}.
In this approach, current is applied to the sample and the difference in the polarization of the incident and reflected light is measured.
The change in the polarization is proportional to the amount of spin magnetic moment that accumulates at the sample edge or surface.
To minimize parasitic thermal effects, the applied current density generally must be kept sufficiently low\cite{riego2016apl,su2017apl,stamm2017prl}.
At such reduced current density, the current-induced MOKE signal in metals is extremely small, typically on the order of a few nanoradians\cite{stamm2017prl,choi2023nature,marui2023prb}.
In contrast, semiconductors exhibit MOKE signals that are several orders of magnitude larger\cite{kato2004science}.
More recently, MOKE measurements have also been used to detect the accumulation of orbital magnetic moments generated by the orbital Hall effect (OHE)\cite{tanaka2008prb,kontani2009prl,go2018prl,jo2018prb,salemi2022prm,choi2023nature,lyalin2023prl,marui2023prb}.

Despite the early discovery of the SHE in semiconductors, the microscopic understanding of the current-induced MOKE signal remains incomplete.
In particular, it is not immediately clear why semiconductors with spin Hall angles much smaller than those of metals can produce significantly larger MOKE signals.
A recent theoretical model\cite{marui2023prb} proposed that the MOKE amplitude depends on several material parameters: the spin Hall angle, resistivity, spin diffusion length, and the energy derivative of the ac spin Hall conductivity at the Fermi energy.
While this model reproduces the magnitude of the MOKE signal observed in transition metals, its general validity across different material classes remains to be tested.

Here we investigate the current-induced MOKE in semimetallic Bi$_{1-x}$Sb$_x$ alloy thin films to address these questions.
Although Bi$_{1-x}$Sb$_x$ is known to host a topological insulator phase in the Bi-rich regime \cite{hsieh2008nature,teo2008prb,zhang2009nphys,khang2018nmat}, we focus on its bulk semi-metallic character by using relatively thick films\cite{fuseya2012jpsj,fuseya2015jpsj,chi2022prb,qu2023prb}.
The transport properties and the MOKE signal are studied as a function of Bi$_{1-x}$Sb$_x$ composition.
We find the MOKE signal scales with the resistivity, carrier density and mobility, which can be accounted for with the model.

Bi$_{1-x}$Sb$_{x}$ thin films were grown on quartz (SiO$_2$) and Si substrates. 
We used the films on quartz substrate to measure the current-induced MOKE and the transport properties. XRD and ellipsometry measurements were performed using the films on Si substrate.
A 1 nm-thick Ta seed layer was first deposited on the substrate using RF magnetron sputtering to promote smooth growth of Bi$_{1-x}$Sb$_{x}$. 
Since the seed layer is predominantly amorphous, we assume the structure of the film is similar for films deposited on quartz and Si substrates.
Bi$_{1-x}$Sb$_{x}$ was then formed using molecular beam epitaxy (MBE) at ambient substrate temperature.
Bi and Sb were co-evaporated to form the alloy.
The temperature of the Knudsen cells was adjusted to tune the deposition rates of Bi and Sb, which define the composition of the film.
The thickness of Bi$_{1-x}$Sb$_{x}$ was set close to 60 nm, significantly larger than the penetration depth of the light (wavelength: 633 nm) used in the experiments.
Finally, a 1 nm-thick Ti capping layer was deposited by electron-beam evaporation to prevent oxidation of the film. 

\begin{figure}[t]
 \begin{center}
  \includegraphics[width=0.9\linewidth]{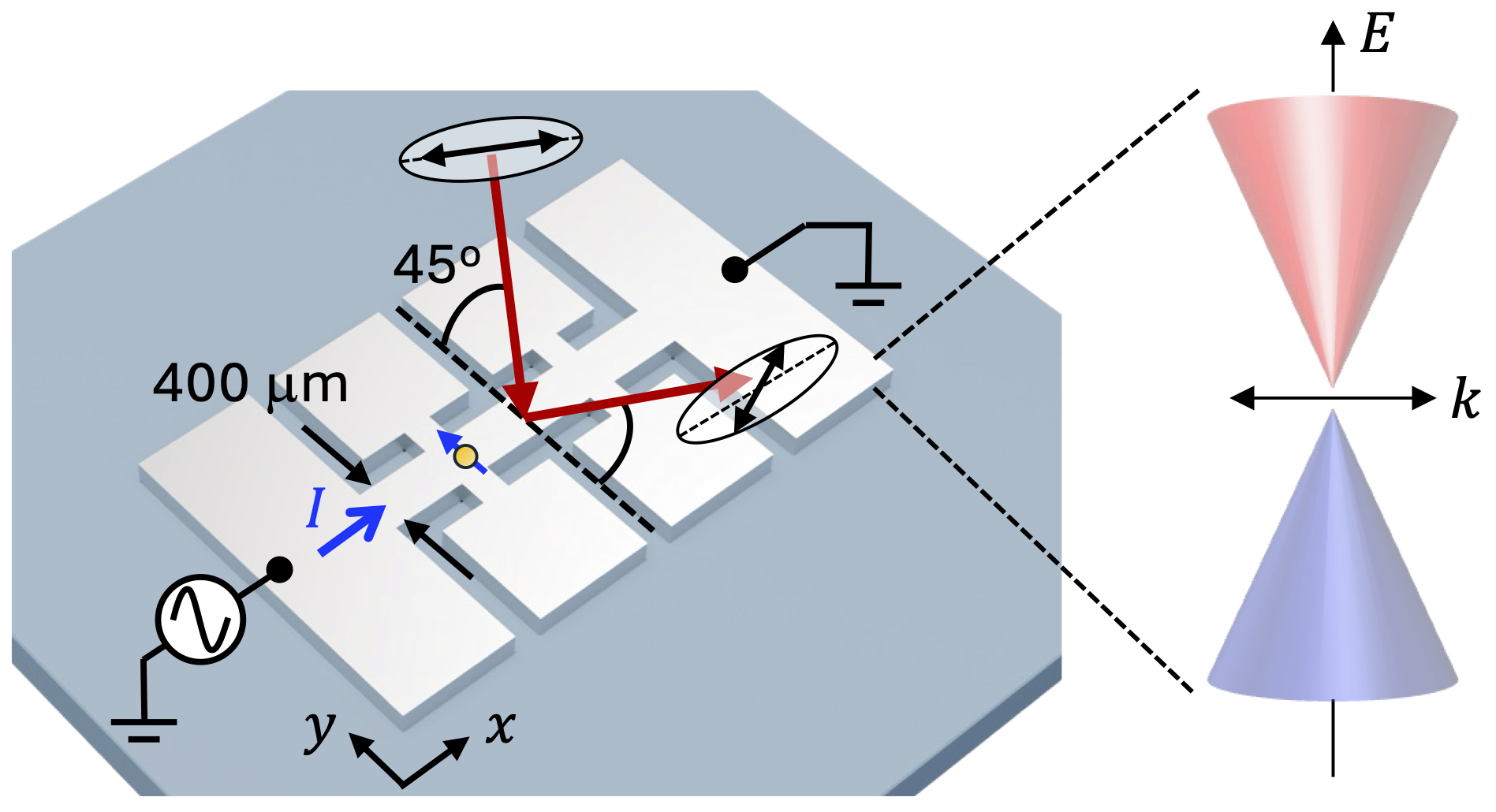}
  \caption{Schematic illustration of the Hall bar, the incident and reflected light, and definition of the coordinate system. As current is passed along the Hall bar, spin Hall effect of Bi$_{1-x}$Sb$_{x}$ induces spin accumulation at the film surface, which is detected by a change in the polarization between the incident and reflected light. The sample is composed of Bi-rich Bi$_{1-x}$Sb$_{x}$, which possess a linear Dirac-like dispersion, as sketched on the right, near the $L-$point in the Briilouin zone. We show the MOKE signal probes the Dirac-like dispersion at the Fermi level.
  }
  \label{fig:summary}
 \end{center}
\end{figure}

Hall bars were made by mechanically scratching the films. 
See Fig.~\ref{fig:summary} for a sketch of the device. 
The width of the current channel of the Hall bars is $\sim$400 $\upmu$m. Electrodes to the Hall bars, made of 2 Ta/10 Al/3 Pt (thickness in nm), were formed using a standard lift-off process. 
The transport properties of the films were characterized using four point probe and Hall resistance measurements.

\begin{figure}[htb]
 \begin{center}
  \includegraphics[width=1.0\linewidth]{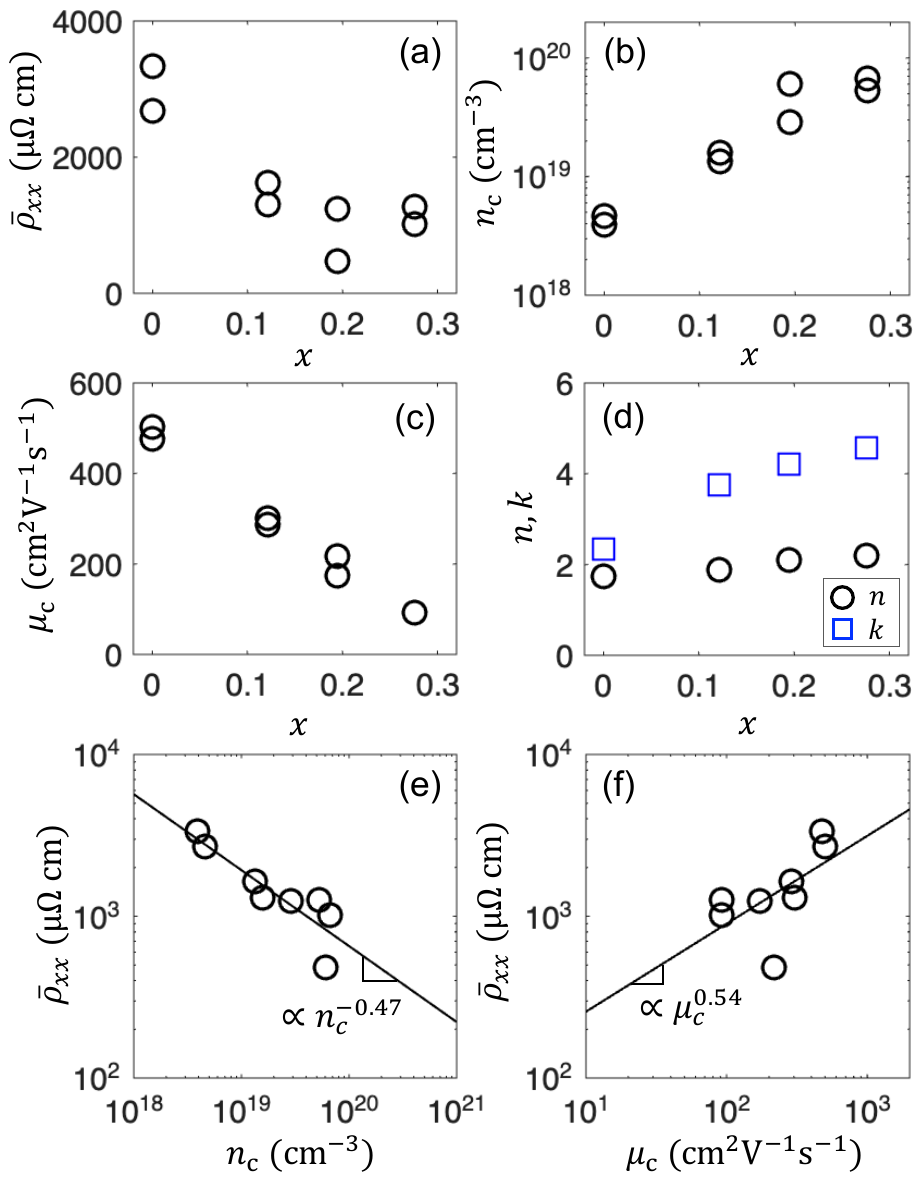}
  \caption{Sb concentration $x$ dependence of the resistivity $\bar{\rho}_{xx}$ (a) carrier density $n_\mathrm{c}$ (b), mobility $\mu_\mathrm{c}$ (c), the refractive index $n$ and the extinction coefficient $k$ (d) of the Bi$_{1-x}$Sb$_{x}$ thin films.   (a-c) Data from two devices on the same substrates are presented. (e,f) $n_\mathrm{c}$ (e) and $\mu_\mathrm{c}$ (f) dependence of $\bar{\rho}_{xx}$ plotted for all films. The solid lines show linear fit to the log-log plot. 
  }
  \label{fig:transport}
 \end{center}
\end{figure}

Figure~\ref{fig:transport}(a) shows the film resistivity $\bar{\rho}_{xx}$ as a function of Sb concentration $x$.
$\bar{\rho}_{xx}$ is obtained by the measured film resistance, the width and length of the Hall bar and the total thickness of the conducting layers (1 nm-thick Ta and 60 nm-thick Bi$_{1-x}$Sb$_{x}$).
Here the difference in the resistivity of the Ta seed and Bi$_{1-x}$Sb$_{x}$ layers is neglected.
Current flow in the 1 nm-thick Ti capping layer, which is heavily oxidized, is neglected.
Overall, $\bar{\rho}_{xx}$ tends to decrease with increasing $x$.
Pure Bi has the largest $\bar{\rho}_{xx}$ which exceeds 2000 $\upmu \Omega$cm.
We use a two carrier model, with equal electron and hole mobilities, to extract the carrier density and mobility from the out of plane magnetic field dependence of the longitudinal and transverse resistances\cite{chi2020sciadv,zhang2024arxiv}.
The $x$ dependence of the carrier density $n_\mathrm{c}$ (the sum of electron and hole densities) and the mobility $\mu_\mathrm{c}$ are plotted in Figs.~\ref{fig:transport}(b) and \ref{fig:transport}(c), respectively.
$n_\mathrm{c}$ increases exponentially with $x$.
In contrast, $\mu_\mathrm{c}$ linearly decreases with $x$.
These trends are consistent with previous reports on Bi$_{1-x}$Sb$_{x}$\cite{vanhulst1995prb,lenoir1996jpcs,chi2020sciadv}.
The change in the resistivity with $x$ is thus primarily caused by the change in $n_\mathrm{c}$.

\begin{figure}[t]
 \begin{center}
  \includegraphics[width=1.0\linewidth]{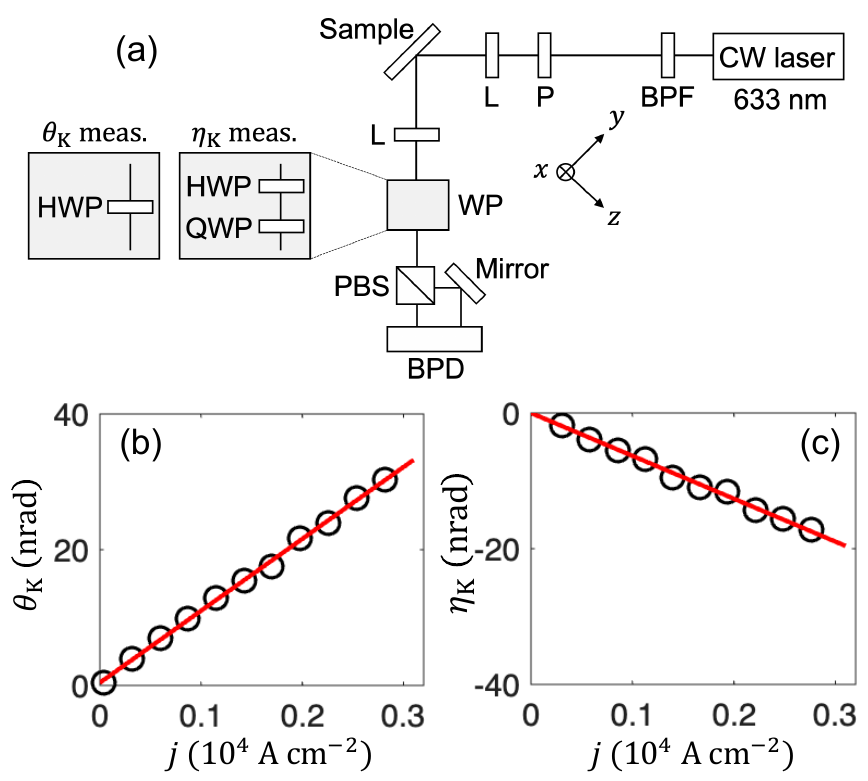}
  \caption{(a) Schematic illustration of the measurement setup and definition of the coordinate system. The incident light is $s$-polarized (polarization is along the $x$-axis). BPF: band-pass filter, P: polarizer, L: lens, HWP: half-wave plate, QWP: quarter-wave plate, PBS: polarized beam splitter, BPD: balanced photodetector. (b,c) Current density $j$ dependence of the real part $\theta_\mathrm{K}$ (b) and the imaginary part $\eta_\mathrm{K}$ (c) of the MOKE signal for pure Bi ($x=0$). The red solid lines show linear fit to the data.  
  }
  \label{fig:jdep}
 \end{center}
\end{figure}

Schematic illustration of the MOKE measurement setup\cite{marui2023prb} is shown in Fig.~\ref{fig:jdep}(a). 
A HeNe laser with a wavelength of 633 nm is used as the light source. 
The beam passes through a band-pass filter and a polarizer to create a linear polarization, and is then directed to the sample from an oblique angle ($\sim$45° from the film normal). 
The diameter of the laser spot at the sample surface is $\sim$25 $\upmu\text{m}$.
The reflected light moves through a half-wave plate (HWP) or combination of a HWP and a quarter-wave plate (QWP) and is split by a polarized beam splitter into two orthogonally polarized components.
These components are detected simultaneously by a balanced photodetector and subsequently converted into electrical signals. 
The signal is fed into a lock-in amplifier synchronized with the frequency ($\sim 2$ kHz) and phase of the ac current applied to the sample.
The real part of the MOKE signal ($\theta_\mathrm{K}$, rotation angle) and the imaginary part ($\eta_\mathrm{K}$, ellipticity angle) were measured using the HWP and HWP/QWP, respectively. 
Positive $\theta_\mathrm{K}$ and $\eta_\mathrm{K}$ is defined as the polarization and elliptical angle of the reflected light rotating clockwise when seen from the sample side.
We vary the ac current amplitude and measure the MOKE signal.
Positive and negative currents indicate ac current that are 180 deg out of phase to each other.
This process is repeated $N$ times ($N \sim  10-100$) to increase the signal to noise ratio.
All measurements were performed at room temperature.

\begin{figure}[tb]
 \begin{center}
  \includegraphics[width=1.0\linewidth]{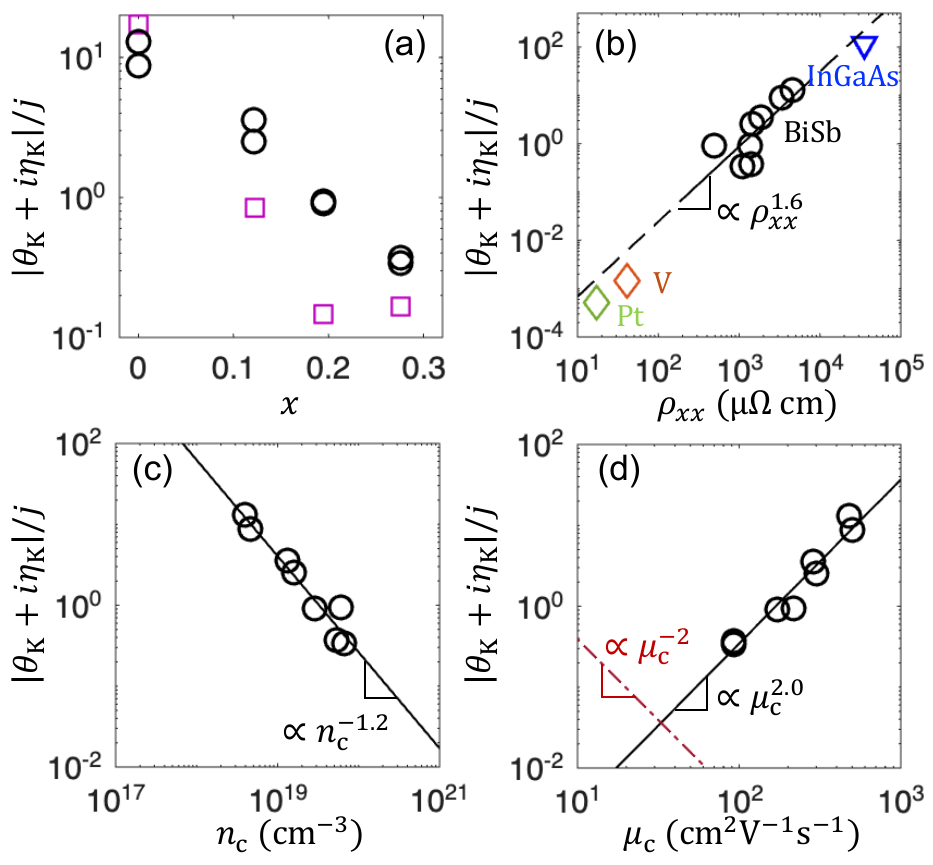}
  \caption{(a) Black circles: Sb concentration $x$ dependence of the magnitude of the MOKE signal per unit current density $|\theta_\mathrm{K} + i \eta_\mathrm{K}| / j$. Purple squares: calculated signal based on Eq.~(\ref{eq:kerr}). Contribution from the orbital Hall effect is neglected.  (b-d) Black circles: $|\theta_\mathrm{K} + i \eta_\mathrm{K}| / j$ of Bi$_{1-x}$Sb$_x$ (BiSb) plotted against the resistivity $\rho_{xx}$ (b), carrier density $n_\mathrm{c}$ (c) and mobility $\mu_\mathrm{c}$ (d). 
  The orange, green diamonds and the blue triangle in (b) show data for V, Pt\cite{marui2023prb} and In$_{0.07}$Ga$_{0.93}$As\cite{kato2004science}, respectively. The black solid lines in (b-d) indicate fit to the data of Bi$_{1-x}$Sb$_x$. The red dashed-dotted line in (d) show calculation result assuming a parabolic free electron model with the intrinsic spin Hall effect and DP-type spin relaxation. (a-d) The unit of the $y$-axis is $10^{-3}$ nrad cm$^2$/A.}
  \label{fig:scaling}
 \end{center}
\end{figure}

Figures~\ref{fig:jdep}(b) and \ref{fig:jdep}(c) show representative current density ($j$) dependence of $\theta_\mathrm{K}$ and $\eta_\mathrm{K}$ for Bi$_{1-x}$Sb$_{x}$ with $x=0$ (pure Bi). 
The current that flows through the film is determined by measuring the voltage across a standard resistor connected in series with the Hall bar. 
The current density $j$ in Bi$_{1-x}$Sb$_{x}$ is estimated by assuming a parallel circuit in which current flows through the two conducting layers, 1 nm-thick Ta and 60 nm-thick Bi$_{1-x}$Sb$_{x}$.
The resistivity of Ta ($\rho_\mathrm{Ta}$) is assumed to be $\sim$200 $\upmu \Omega$cm, obtained from experiments on similar systems\cite{kim2016prl}.
We also estimate the resistivity of Bi$_{1-x}$Sb$_{x}$, defined as $\rho_{xx}$, from the film resistivity $\bar{\rho}_{xx}$, $\rho_\mathrm{Ta}$ and the layer thicknesses, using the parallel circuit model. 
Both $\theta_\mathrm{K}$ and $\eta_\mathrm{K}$ linearly increase with $j$, suggesting that effects due to current-induced heating is negligible.
We fit the data with a linear function to extract the rate at which the MOKE signal changes with $j$, i.e. $\theta_\mathrm{K} / j$ and $\eta_\mathrm{K} / j$. 
The magnitude of the MOKE signal, $| \theta_\mathrm{K} + i \eta_\mathrm{K} | / j$ is plotted against $x$ in Fig.~\ref{fig:scaling}(a), black circles.
As is evident, the magnitude of the MOKE signal monotonically decreases with $x$ in log scale.

A phenomenological model was previously developed\cite{marui2023prb} to account for the MOKE signal.
According to this model, $\theta_\mathrm{K}$ and $\eta_\mathrm{K}$ are expressed as the following:
\begin{equation}
\begin{aligned}
\label{eq:kerr}
\theta_\mathrm{K} + i \eta_\mathrm{K} = - \left. \frac{i e}{\varepsilon_0 \omega} \frac{\partial \sigma_\mathrm{s} (\omega)}{\partial E} \right|_{E_\mathrm{F}} l_\mathrm{s} \sigma_\mathrm{s}(0) \rho_{xx}^2 j h(\varepsilon_{xx}, \theta, t, l_\mathrm{s}). 
\end{aligned}
\end{equation}
$\sigma_\mathrm{s} (0)$ and $l_\mathrm{s}$ are the dc spin Hall conductivity and the spin diffusion length, respectively.
$\sigma_\mathrm{s} (\omega)$ is the ac spin Hall conductivity driven by the light electric field with angular frequency $\omega$.
$\left. \partial \sigma_\mathrm{s} (\omega) / \partial E \right|_{E_\mathrm{F}}$ is the energy derivative of $\sigma_\mathrm{s} (\omega)$ evaluated at the Fermi energy $E_\mathrm{F}$.
$\left. \partial \sigma_\mathrm{s} (\omega) / \partial E \right|_{E_\mathrm{F}}$ is equivalent to the derivative of the transverse conductivity with (spin dependent) carrier density, where the latter is known to represent inter-band transitions\cite{marui2023prb}.
$h(\varepsilon_{xx}, \theta, t, l_\mathrm{s})$ is a complex function of the diagonal component of the permittivity tensor $\varepsilon_{xx}$, the angle $\theta$ between the incident light wave vector and the film normal, film thickness $t$ and $l_\mathrm{s}$.
See Appendix, Sec.~\ref{sec:app:model} for the exact form of $h(\varepsilon_{xx}, \theta, t, l_\mathrm{s})$.
$\varepsilon_0$ and $e$ are the vacuum permittivity and the electric charge, respectively.

First, we show that Eq.~(\ref{eq:kerr}) provides order of magnitude estimates of the MOKE signal that are consistent with the experiments. 
Here we neglect contribution from the orbital Hall effect for simplicity: see Appendix, Sec.~\ref{sec:app:model}, for the calculation results when the orbital Hall effect is considered.
The Maxwell's equations are solved numerically\cite{yeh1980surfsci,visnovsky1986czechjpb,zak1990jmmm} to compute $h(\varepsilon_{xx}, \theta, t, l_\mathrm{s})$\cite{marui2023prb}.
The dc and ac spin Hall conductivities are obtained from tight binding model calculations of the electronic structure of Bi$_{1-x}$Sb$_{x}$ and the Kubo formula\cite{liu1995prb,sahin2015prl,qu2023prb}.
The $x$-dependence of $\sigma_\mathrm{s}(0)$ and $\left. \partial \sigma_\mathrm{s}(\omega) / \partial E \right|_{E_\mathrm{F}}$ are shown in the Appendix, Fig.~\ref{fig:modelcal}(d-f). 
$\sigma_\mathrm{s}(0)$ is nearly constant in the range of $x$ shown here, consistent with previous calculations\cite{sahin2015prl}.
Values of $\rho_{xx}$ extracted from the film resistivity $\bar{\rho}_{xx}$ are shown in the Appendix, Table~\ref{table:resistivity}.
Here we take the average of $\rho_{xx}$ obtained from two devices for a given $x$.
$\varepsilon_{xx}$ is determined by the refractive index $n$ and the extinction coefficient $k$, obtained using standard ellipsometry measurements.
The $x$ dependence of $n$ and $k$ are shown in Fig.~\ref{fig:transport}(d).
$n$ and $k$ tend to increase with increasing $x$.
$l_\mathrm{s}$ for Bi ($\sim18$ nm) was estimated previously\cite{hirose2021prb}.
We set $l_\mathrm{s}\sim18$ nm for all films.
Such assumption is consistent with Dyakonov-Perel (DP)-type spin relaxation, where $l_\mathrm{s}$ shows little dependence on the resistivity: see the discussion pertaining to Eq.~(\ref{eq:diffusionconst}) for the details.

The purple squares in Fig.~\ref{fig:scaling}(a) show the $x$ dependence of the calculated MOKE signal ($| \theta_\mathrm{K} + i \eta_\mathrm{K} | / j$) using Eq.~(\ref{eq:kerr}).
As is evident, the calculation roughly reproduces the trend of the MOKE signal as a function of $x$.
Whereas the calculated results are close to that of the experiments for $x=0$, the former is overall smaller than that of the latter when $x>0$.
Including the contribution from the orbital Hall effect mitigates the difference: see Appendix, Sec.~\ref{sec:app:model}. 
It is also possible that the topological surface states, which is not included in the model calculations, contribute to the MOKE signal when $0.09 \lesssim x \lesssim 0.22$\cite{sahin2015prl}.

Next, we study the relations between the transport properties and the MOKE signal to further test the validity of Eq.~(\ref{eq:kerr}).
In Fig.~\ref{fig:scaling}(b-d), we show $|\theta_\mathrm{K} + i \eta_\mathrm{K}| / j$ plotted against $\rho_{xx}$, $n_\mathrm{c}$ and $\mu_\mathrm{c}$ using a log-log plot.
$|\theta_\mathrm{K} + i \eta_\mathrm{K}| / j$ scales with $\rho_{xx}$, $n_\mathrm{c}$ and $\mu_\mathrm{c}$.
The log-log plot is fitted with a linear function to obtain the exponent of the scaling relations.
The fitting results, shown by the black solid lines, agree well with the data.
In Fig.~\ref{fig:scaling}(b), data for In$_{0.07}$Ga$_{0.93}$As, Pt, V, extracted from Refs.~\cite{kato2004science} and \cite{marui2023prb}, are presented together. 
Overall, we find that the difference in the MOKE signal can be roughly accounted for with the difference in $\rho_{xx}$.
In the following, we compare the exponents obtained from the fitting with those predicted by Eq.~(\ref{eq:kerr}). 

As some of the parameters in Eq.~(\ref{eq:kerr}) are implicitly dependent on $\rho_{xx}$, $n_\mathrm{c}$, $\mu_\mathrm{c}$, here we clarify the relations. 
For this purpose, it is convenient to introduce the relaxation time $\tau_\mathrm{eff}$ which determines the carrier transport properties.
Let us assume that $\tau_\mathrm{eff}$ satisfies the following general relation\cite{dassarma2013prb,dassarma2015prb,chi2022prb}
\begin{equation}
\begin{aligned}
\tau_\mathrm{eff} = a_0 n_\mathrm{c}^{\alpha},
\label{eq:taueff}
\end{aligned}
\end{equation}
where $a_0$ and $\alpha$ are constants.
Previous studies reported that $\alpha$ depends on the electronic structure of the host material.
For example, $\alpha = - \frac{1}{3}$ when the band structure is described by the Dirac Hamiltonian\cite{dassarma2015prb,chi2022prb} (we refer to electrons in such system as Dirac electrons).
For free electrons in a parabolic band, $\alpha = 0$ is often adopted.
Bi-rich Bi$_{1-x}$Sb$_{x}$ alloys used in this study belong to the former: the carriers near the L point of the Brillouin zone at the Fermi level are categorized as Dirac electrons\cite{sahin2015prl,chi2020sciadv}.
Although we cannot distinguish effects from topological surface states\cite{teo2008prb,zhang2009nphys,hsieh2008nature,benia2015prb,khang2018nmat,baringthon2022prm}, if any, for simplicity here we focus on spin transport properties that originate from the bulk states and thus compare model calculation results assuming Dirac electrons and free electrons.
The value of $\alpha$ for the Dirac model is determined using the experimental results presented in Fig.~\ref{fig:transport}, while we fix $\alpha = 0$ for the free electron model.
\begin{table}[b]
    \caption{
    Values of the exponent $\nu$ for the scaling relations $\rho_{xx} \propto \tau_\mathrm{eff}^{\nu}$, $\rho_{xx} \propto n_\mathrm{c}^{\nu}$ and $\rho_{xx} \propto \mu_\mathrm{c}^{\nu}$.
    }
    \label{table:scaling:res}
    \centering
    \begin{tabular}{c c c c}
    \toprule
   model & $\tau_\mathrm{eff}^{\nu}$ & $n_\mathrm{c}^{\nu}$ & $\mu_\mathrm{c}^{\nu}$\\
    \midrule
   Dirac & $-1 - \frac{2}{3 \alpha}$ & $-\alpha -\frac{2}{3}$ & $-\frac{3 \alpha + 2}{3 \alpha - 1}$ \\
   Free & $-1$ & $-1$ & $-1$ \\
   Exp. & n/a & $-0.47$ & $0.54$\\
   \bottomrule
    \end{tabular}
\end{table}

In Fig.~\ref{fig:transport}(e,f), we show the relation between $\rho_{xx}$ and $n_\mathrm{c}$, $\mu_\mathrm{c}$. 
The resistivity decreases with increasing $n_\mathrm{c}$, however, it increases with increasing $\mu_\mathrm{c}$.
This is in stark contrast with metals, where $\rho_{xx}$ typically decreases with increasing $\mu_\mathrm{c}$.
As we show below, these scalings reflect the character of Dirac electron system.
We fit the log-log plot in Fig.~\ref{fig:transport}(e,f) with a linear line to obtain the scaling exponents.
To obtain the scaling relation, we first express $\rho_{xx} = \left( n_\mathrm{c} e \mu_\mathrm{c} \right)^{-1}$ in terms of $\tau_\mathrm{eff}$.
For Dirac electrons, $\mu_\mathrm{c}$ possesses a unique relation with $n_\mathrm{c}$\cite{chi2022prb}, which reads
\begin{equation}
\begin{aligned}
\mu_\mathrm{c} = a_1 n_\mathrm{c}^{\alpha - 1/3},
\label{eq:mobility:density:dirac}
\end{aligned}
\end{equation}
where $a_1$ is a constant. The mobility of the Dirac and free electron systems is therefore given by
\begin{equation}
\mu_\mathrm{c} = 
\begin{aligned}
\begin{cases}
\left( a_1 a_0^{-1+ \frac{1}{3\alpha}} \right) \tau_\mathrm{eff}^{1 - \frac{1}{3\alpha} }: \ \mathrm{Dirac},\\
\left( \frac{e}{m_\mathrm{eff}} \right) \tau_\mathrm{eff}: \ \mathrm{Free},
\end{cases}
\label{eq:mobility:tau}
\end{aligned}
\end{equation}
where the former is derived from Eqs.~(\ref{eq:taueff}) and (\ref{eq:mobility:density:dirac}), and the latter is based on the Drude model.
Substituting Eqs.~(\ref{eq:taueff}) and (\ref{eq:mobility:tau}) into $\rho_{xx} = \left( n_\mathrm{c} e \mu_\mathrm{c} \right)^{-1}$, we obtain the the scaling relations between $\rho_{xx}$ and $\tau_\mathrm{eff}$, from which we find the relations between $\rho_{xx}$ and $n_\mathrm{c}$, $\mu_\mathrm{c}$.
The relations are summarized in Table~\ref{table:scaling:res}.
For the free electron model, the exponents are all -1, consistent with the Drude model.
The exponents derived from the Dirac model depends on $\alpha$. 
For example, if we substitute $\alpha = - \frac{1}{3}$, we find $\rho_{xx} \propto n_\mathrm{c}^{-1}$ and $\rho_{xx} \propto \mu_\mathrm{c}^{\frac{1}{2}}$.
As noted above, in the Dirac model, $\rho_{xx}$ increases with increasing $\mu_\mathrm{c}$.

In Table~\ref{table:scaling:res}, the exponents obtained from the experiments are presented in the bottom row.
We equate these exponents with the form from the Dirac model to determine $\alpha$: we find $\alpha \sim -0.20$ and $\alpha \sim -0.32$  from the scaling of $\rho_{xx}$ with $n_\mathrm{c}$ and $\mu_\mathrm{c}$, respectively.
These values are close to what the theory predicts for the Dirac electrons: $\alpha = - \frac{1}{3}$\cite{dassarma2015prb}.
For the Dirac model, we therefore assume that $\alpha$ lies in between $-0.20$ and $-0.32$.
\begin{table}[b]
    \caption{Forms of the exponent $\nu$ for the scaling relations $|\theta_\mathrm{K} + i \eta_\mathrm{K}| \propto \tau_\mathrm{eff}^{\nu}$, $|\theta_\mathrm{K} + i \eta_\mathrm{K}| \propto n_\mathrm{c}^{\nu}$, $|\theta_\mathrm{K} + i \eta_\mathrm{K}| \propto \mu_\mathrm{c}^{\nu}$ and $|\theta_\mathrm{K} + i \eta_\mathrm{K}| \propto \rho_{xx}^{\nu}$. $A \equiv \beta + \frac{1}{2}(1 + \delta) - 2 \frac{3 \alpha + 2}{3 \alpha}$ and $B \equiv \beta + \frac{1}{2}(1 + \delta) - 2$.
    }
    \label{table:scaling:MOKE}
    \centering
    \begin{tabular}{c c c c c}
    \toprule
   model & $\tau_\mathrm{eff}^{\nu}$ & $n_\mathrm{c}^{\nu}$ & $\mu_\mathrm{c}^{\nu}$ & $\rho_{xx}^{\nu}$\\
    \midrule
   Dirac & $A$ & $\alpha A$ & $\frac{3 \alpha}{3 \alpha - 1} A $ & $- \frac{3 \alpha}{3 \alpha + 2} A $ \\
   Free & $B$ & $0$ & $B$ & $-B$ \\
   \bottomrule
    \end{tabular}
\end{table}

We now examine the relation between the MOKE amplitude $|\theta_\mathrm{K} + i \eta_\mathrm{K}|$  and $n_\mathrm{c}$, $\mu_\mathrm{c}$, $\rho_{xx}$ based on Eq.~(\ref{eq:kerr}).
In doing so, we must determine how $l_\mathrm{s}$ and $\sigma_\mathrm{s}$ depend on $\tau_\mathrm{eff}$.
The spin diffusion length is given by $l_\mathrm{s} = \sqrt{D \tau_\mathrm{s}}$, where the diffusion constant $D$ for the Dirac electrons\cite{burkov2011prl} and the free electrons are given by 
\begin{equation}
\begin{aligned}
D = 
\begin{cases}
\frac{1}{3} v_\mathrm{F}^2 \tau_\mathrm{eff}: \ \ \textrm{Dirac,}\\
\frac{E_\mathrm{F}}{e} \mu_\mathrm{c}  = \frac{E_\mathrm{F}}{m_\mathrm{eff}} \tau_\mathrm{eff}: \ \ \textrm{Free}.
\end{cases}
\label{eq:diffusionconst}
\end{aligned}
\end{equation}
$v_\mathrm{F}$ is the Fermi velocity and $m_\mathrm{eff}$ is the effective mass.
For the free electrons, we used the Einstein relation and Eq.~(\ref{eq:mobility:tau}).
$\tau_\mathrm{s}$ is the spin relaxation time, given by 
\begin{equation}
\begin{aligned}
\tau_\mathrm{s} = a_2 \tau_\mathrm{eff}^\delta,
\label{eq:spinrelxationtime}
\end{aligned}
\end{equation}
where $a_2$ is a constant, $\delta = -1$ $[1]$ for the Dyakonov-Perel (DP) [Elliott-Yafet (EY)]-type spin relaxation.
For the dc spin Hall conductivity $\sigma_\mathrm{s}(0)$, we consider two contributions, intrinsic and extrinsic (skew scattering). It is common to assume 
\begin{equation}
\begin{aligned}
\sigma_\mathrm{s}(0) = a_3 \tau_\mathrm{eff}^\beta.
\label{eq:spinhallconductivity}
\end{aligned}
\end{equation}
where $a_3$ is a constant.
$\beta = 0$ for the intrinsic SHE and $\beta = 1$ for the extrinsic SHE (skew scattering).
Note that such relation and those shown in Table~\ref{table:scaling:res} return $\rho_\mathrm{s}(0) \propto \rho_{xx}^{-\beta + 2}$ for the free electron model, where $\rho_\mathrm{s}(0)$ is the dc spin Hall resistivity.
The scaling between $\rho_\mathrm{s}(0)$ and $\rho_{xx}$ has been used to determine the origin of the spin Hall effect in metals\cite{vila2007prl,sinova2015rmp}.
We assume the ac spin Hall conductivity $\sigma_\mathrm{s}(\omega)$ is not dependent on $\tau_\mathrm{eff}$ as the relaxation time is typically significantly larger than the period of the light electric field. 
For example, in pure Bi, $\tau_\mathrm{eff} \sim 10^{-12} - 10^{-11}$ s\cite{chi2022prb}, whereas the period of the light is $2 \times 10^{-15}$ s.
With these relations, the variation of $|\theta_\mathrm{K} + i \eta_\mathrm{K}|$ with $n_\mathrm{c}$, $\mu_\mathrm{c}$ and $\rho_{xx}$ can be derived, which are summarized in Table~\ref{table:scaling:MOKE}.
\begin{figure}[b]
 \begin{center}
  \includegraphics[width=1.0\linewidth]{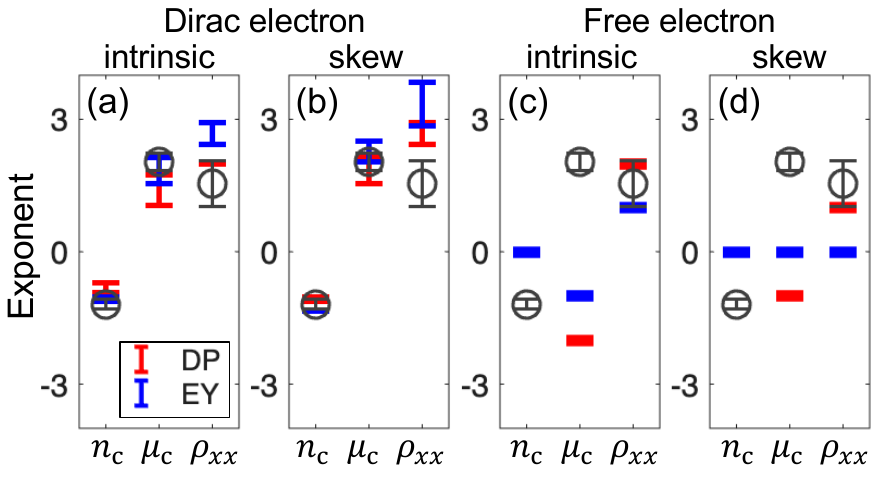}
  \caption{(a-d) Values of the exponent $\nu$ for the scaling relations $|\theta_\mathrm{K} + i \eta_\mathrm{K}| \propto n_\mathrm{c}^{\nu}$ (left), $\mu_\mathrm{c}^{\nu}$ (center) and $\rho_{xx}^{\nu}$ (right). 
    The open circles display the $\nu$ obtained in the experiments. The attached black error bars represent the fitting error when a linear function is fitted to the log-log plot shown in Fig.~\ref{fig:scaling}(b-d).
    The horizontal colored bars shows the $\nu$ estimated using the Dirac model (a,b) and the free electron model (c,d) with DP (red bars, $\delta = -1$) and EY (blue bars, $\delta = 1$) spin relaxations. (a,c) and (b,d) show results with the intrinsic ($\beta = 0$) and extrinsic ($\beta = 1$) SHE. The range of the horizontal bar for the Dirac model represents calculation results with $-0.32 \leq \alpha  \leq -0.20$. $\alpha = 0$ for the free electron model.
}
 \label{fig:exponents}
 \end{center}
\end{figure}

We calculate the exponents presented in Table~\ref{table:scaling:MOKE} by substituting the appropriate values of $\alpha$, $\beta$ and $\delta$.
For the free electron model with an intrinsic SHE and a DP-type spin relaxation, we find $|\theta_\mathrm{K} + i \eta_\mathrm{K}| \propto n_\mathrm{c}^{0}$, $|\theta_\mathrm{K} + i \eta_\mathrm{K}| \propto \rho_{xx}^{2}$ and $|\theta_\mathrm{K} + i \eta_\mathrm{K}| \propto \mu_\mathrm{c}^{-2}$.
The last relation is drawn by the red dash-dotted line in Fig.~\ref{fig:scaling}(d) for reference.
The scaling between $|\theta_\mathrm{K} + i \eta_\mathrm{K}|$ and $\rho_{xx}$ roughly agrees with the trend found in the experiments.
However, the scaling relation of $|\theta_\mathrm{K} + i \eta_\mathrm{K}|$ with $\mu_\mathrm{c}$ is clearly inconsistent with the experimental results of Bi$_{1-x}$Sb$_{x}$.

To illustrate how the scaling relations differ between the Dirac and free electron models in more detail, we show the calculated exponents for $|\theta_\mathrm{K} + i \eta_\mathrm{K}|$ vs. $n_\mathrm{c}$, $\mu_\mathrm{c}$ and $\mu_\mathrm{c}$ in Fig.~\ref{fig:exponents}(a,b) [Dirac electrons] and \ref{fig:exponents}(c,d) [free electrons].
Figure~\ref{fig:exponents}(a,c) [(b,d)] show the results when the intrinsic (extrinsic) SHE is considered. 
The red and blue bars represent results when the DP-type and EY-type spin relaxations are assumed.
The range of the colored vertical bars for the Dirac electrons [Fig.~\ref{fig:exponents}(a,b)] indicate variation of the exponents when $\alpha$ is varied from -0.32 to -0.2.
For the free electron model, $\alpha$ is fixed to 0, thus the symbol is a single line.
The exponents obtained from the experiments (Fig.~\ref{fig:scaling}) are shown using the open circles.
We find that the Dirac model describes the experimental results of Bi$_{1-x}$Sb$_{x}$ better than the free electron model.
This is primarily because the experiments dictate the exponents of $|\theta_\mathrm{K} + i \eta_\mathrm{K}|$ vs. $\mu_\mathrm{c}$ and $\rho_{xx}$ have the same sign, which is not the case for the free electron model.
From the comparison presented in Fig.~\ref{fig:exponents}, it is difficult to determine the origin of the SHE (intrinsic vs. extrinsic) or the spin relaxation mechanism (DP vs. EY) .
The calculation results do not significantly change if one assumes the ac spin Hall conductivity to be an extrinsic origin, i.e. $\sigma_\mathrm{s}(\omega) \propto \tau_\mathrm{eff}$.
Recent studies have reported an unconventional scaling of the orbital Hall conductivity $\sigma_\mathrm{o}(0)$ with the resistivity, that is, assuming a free electron model, $\sigma_\mathrm{o}(0) \propto \rho_{xx} \propto \tau_\mathrm{eff}^{-1}$\cite{peng2025ncomm}.
However, we find such scaling cannot account for the experimental results.

These results therefore show that the model developed to describe current induced MOKE in metals can also account for the results from semi-metallic Bi$_{1-x}$Sb$_{x}$.
Moreover, the MOKE signal scales with the transport parameters that reflect the Dirac-like band structure of Bi$_{1-x}$Sb$_{x}$\cite{li2008science,zhu2011prb,teo2008prb,sahin2015prl,chi2022prb,fukushima2024prb}.
Finally, recent studies have suggested that the current induced MOKE signal may contain contribution from electro-optic effects\cite{konig2019prb,mahfouzi2025prb1,mahfouzi2025prb2,tanabe2025jjap}.
We find that the MOKE signal does not change its sign when we change the incident light polarization from $s$ to $p$, which indicates that contribution from the intrinsic electro-optic effect is small\cite{mahfouzi2025prb2}.
Further studies are required to clarify the contributions from the extrinsic electro-optic effect (e.g. current induced orbital Kerr effect)\cite{ovalle2024prb}. 

In summary, we have investigated the current-induced MOKE in semi-metallic Bi$_{1-x}$Sb$_x$ alloy thin films. 
The MOKE signal increases with decreasing Sb concentration, reaching its maximum in pure Bi ($x=0$), where it exceeds that of typical transition metals by nearly four orders of magnitude. 
The signal scales with the resistivity as $\rho_{xx}^{1.7 \pm 0.6}$ and with the carrier mobility as $\mu_\mathrm{c}^{2.0 \pm 0.2}$.
These scaling relations are consistent with model calculations that assume the Dirac electrons are responsible for the generation and accumulation of spin magnetic moment, in contrast to the $\rho_{xx}^{2}$ and $\mu_\mathrm{c}^{-2}$ scaling predicted by the parabolic free electron model.
Although the scaling exponent of MOKE with resistivity is different between the free electron-like transition metals and the Dirac semimetals, the scaling with $\rho_{xx}$ can mostly account for the order of magnitude differences found among metals, semi-metals and semiconductors. 
These results thus demonstrate that current-induced MOKE serves as a sensitive probe of spin current and the underlying electronic state in a broad class of materials.

\begin{acknowledgments}
We thank ARIM for the support on the ellipsometry measurements. This work was partly supported by JSPS KAKENHI (Grant Number 23H00176), JST CREST (JPMJCR19T3), MEXT Initiative to Establish Next-generation Novel Integrated Circuits Centers (X-NICS) and Cooperative Research Project Program of RIEC, Tohoku University. R.M. thanks Support Center for Advanced Telecommunications Technology Research (SCAT) and S.W. thanks JST SPRING GX, Grant Number JPMJSP2108.
\end{acknowledgments}

\section{\label{sec:app} Appendix}
\subsection{\label{sec:app:moke} Transport properties}
In Table~\ref{table:resistivity}, we show the estimated resistivity of Bi$_{1-x}$Sb$_{x}$ ($\rho_{xx}$) using the parallel circuit model and the film resistivity ($\bar{\rho}_{xx}$) [see Fig.~\ref{fig:transport}(a)].
We take the average value of the two devices measured for a given composition $x$.
The resistivity of the 1 nm-thick Ta seed layer is assumed to be 200 $\upmu\Omega$cm\cite{kim2016prl}.
\begin{table}[h]
    \caption{Average resistivity of Bi$_{1-x}$Sb$_{x}$ ($\rho_{xx}$) estimated from the parallel circuit model and the film resistivity ($\bar{\rho}_{xx}$).}
    \label{table:resistivity}
    \centering
    \begin{tabular}{c c}
    \toprule
   $x$ & $\rho_{xx}$ ($\upmu \Omega$ cm)\\
    \midrule
   0 & 3913\\
   0.12 & 1644\\
   0.20 & 910\\
   0.28 & 1237\\
   \bottomrule
   \end{tabular}
\end{table}

\subsection{\label{sec:app:moke} MOKE signal}
\begin{figure}[h]
 \begin{center}
  \includegraphics[width=1\linewidth]{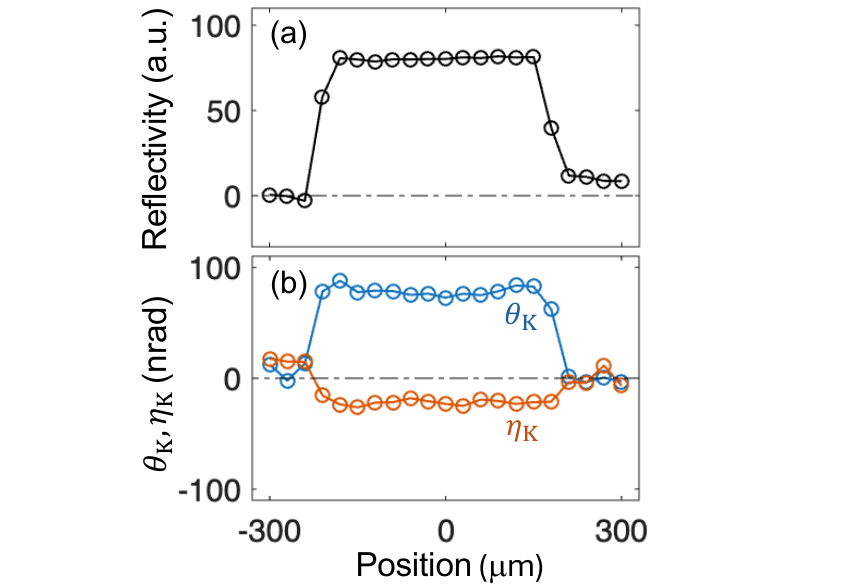}
  \caption{(a,b) Line profile of the reflectivity (a) and $\theta_\mathrm{K}$ and $\eta_\mathrm{K}$ (b) along the wire width (orthogonal to the current flow) for Bi$_{1-x}$Sb$_{x}$ with $x= 0$.
  }
  \label{fig:lineprofile}
 \end{center}
\end{figure}

In Fig.~\ref{fig:lineprofile}(a), we show a line profile of light reflectivity along the wire width (orthogonal to the current flow) for Bi$_{1-x}$Sb$_{x}$ with $x= 0$. The corresponding MOKE signals ($\theta_\mathrm{K}$ and $\eta_\mathrm{K}$) are shown in Fig.~\ref{fig:lineprofile}(b). We find negligible contribution from the out of plane spin magnetic moment induced by the Oersted field, likely due to the small current passed along the wire.

$\theta_\mathrm{K} / j$ and $\eta_\mathrm{K} / j$ are plotted against the Sb concentration $x$ in Figs.~\ref{fig:modelcal}(a) and \ref{fig:modelcal}(b), respectively, using open circles.
Both $\theta_\mathrm{K} / j$ and $\eta_\mathrm{K} / j$ decrease with increasing $x$.
The $x$ dependence of $| \theta_\mathrm{K} + i \eta_\mathrm{K} | / j$ is plotted in Fig.~\ref{fig:modelcal}(c): the data (open circles) are the same with those presented in Fig.~\ref{fig:scaling}(a).

\subsection{\label{sec:app:model} Model calculations}
Here, we show the calculation results when contribution from the orbital Hall effect on the MOKE signal is included.
Following the approach described in Ref.~\cite{marui2023prb}, $\theta_\mathrm{K}$ and $\eta_\mathrm{K}$ are expressed as the following:
\begin{equation}
\begin{aligned}
\label{eq:kerr:so}
\theta_\mathrm{K} + i \eta_\mathrm{K} \sim \langle \left( \varepsilon_{zx}^\mathrm{s} + \varepsilon_{zx}^\mathrm{o} \right) g(\varepsilon_{xx}, \theta) \rangle. 
\end{aligned}
\end{equation}
$g(\varepsilon_{xx}, \theta)$ represents factors related to the reflection of light from the film under an oblique incidence.
$\langle X \rangle$ indicates $X$ integrated along the film thickness.
The off-diagonal component of the permittivity tensor $\varepsilon_{zx}^{\mathrm{s[o]}}$ is calculated using the following relation\cite{marui2023prb}:
\begin{equation}
\begin{aligned}
\varepsilon_{zx}^\mathrm{s[o]} = \left. \frac{i}{\varepsilon_0 \omega} \frac{n_\mathrm{s[o]}}{D_\mathrm{F}} \frac{\partial \sigma_\mathrm{s[o]} (\omega)}{\partial E} \right|_{E_\mathrm{F}}
\label{eq:permittivity}
\end{aligned}
\end{equation}
where the spin [orbital] accumulation $n_\mathrm{s[o]}$ is obtained by solving the spin [orbital] diffusion equation and reads
\begin{equation}
\begin{aligned}
n_\mathrm{s[o]} = - e D_\mathrm{F} l_\mathrm{s[o]} \sigma_\mathrm{s[o]}(0) \rho_{xx}^2 j_x \frac{\sinh \left(\frac{t-2z}{2 l_\mathrm{s[o]}} \right)}{\cosh \left(\frac{t}{2 l_\mathrm{s[o]}} \right)}.
\label{eq:spinaccum}
\end{aligned}
\end{equation}
$\sigma_\mathrm{s[o]} (0)$ and $l_\mathrm{s[o]}$ are the dc spin [orbital] Hall conductivity and the associated diffusion length, respectively.
$\sigma_\mathrm{s[o]} (\omega)$ is the ac spin [orbital] Hall conductivity driven by the light electric field with angular frequency $\omega$.
$\left. \partial \sigma_\mathrm{s[o]} (\omega) / \partial E \right|_{E_\mathrm{F}}$ is the energy derivative of $\sigma_\mathrm{s[o]} (\omega)$ evaluated at the Fermi energy.
$D_\mathrm{F}$ is the density of states at the Fermi energy.
$z$ is a coordinate within the film, where $z = 0$ and $z = t$ corresponds to the bottom and top surfaces of the film that face the substrate and air, respectively. 
\begin{figure}[tb]
 \begin{center}
  \includegraphics[width=1\linewidth]{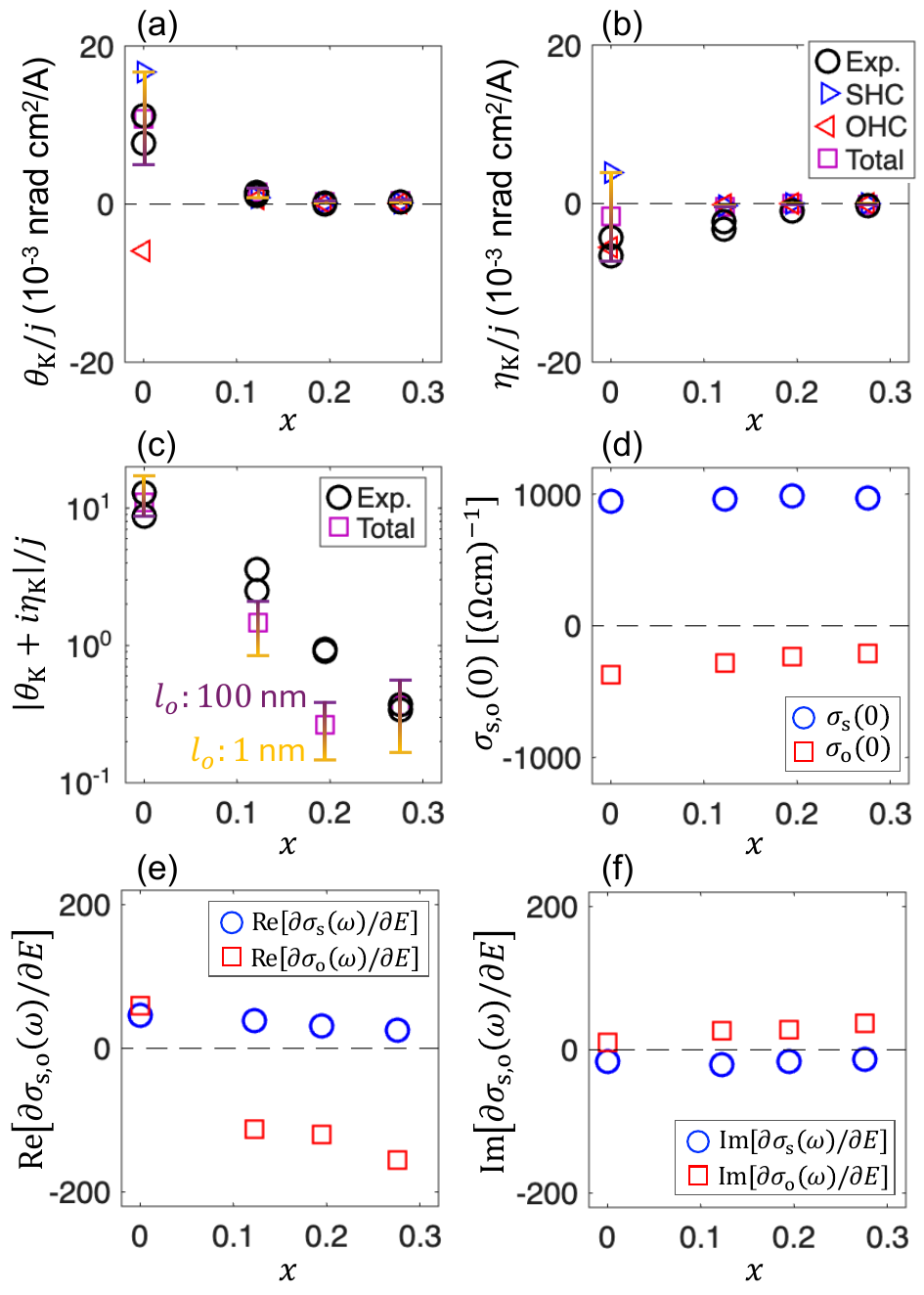}
  \caption{Sb concentration $x$ dependence of the real part $\theta_\mathrm{K}$ (a), the imaginary part $\eta_\mathrm{K}$ (b) and the magnitude $|\theta_\mathrm{K} + i \eta_\mathrm{K}|$ (c) of current induced MOKE signal normalized by the current density $j$. Black circles represent experimental data. Data from two devices on the same substrates are presented. Blue and red triangles and purple squares show contributions from the spin Hall effect, the orbital Hall effect and the sum of the two, respectively. 
  The symbols show results when $l_\mathrm{s} = l_\mathrm{o} = 18$ nm for all $x$. The yellow and purple error bars attached to the squares in (a-c) indicate the calculated MOKE signal when $l_\mathrm{o}$ is set to 1 nm and 100 nm, respectively.   The unit of the $y$-axis of (c) is $10^{-3}$ nrad cm$^2$/A. 
  (d) Sb concentration $x$ dependence of dc spin Hall conductivity $\sigma_\mathrm{SH} (0)$ and dc orbital Hall conductivity $\sigma_\mathrm{OH} (0)$. (e,f) Real (e) and imaginary (f) parts of $\left. \frac{\partial \sigma_\mathrm{s[o]} (\omega)}{\partial E} \right|_{E_\mathrm{F}}$ plotted against $x$.
The unit of the $y$-axis of (e,f) is $(\Omega \ \mathrm{cm \ eV})^{-1}$. 
  }
  \label{fig:modelcal}
 \end{center}
\end{figure}

The Maxwell's equations are solved numerically to compute the thickness averaged value of $\theta_\mathrm{K} + i \eta_\mathrm{K}$\cite{yeh1980surfsci,visnovsky1986czechjpb,zak1990jmmm,marui2023prb}.
When solving the equations, material parameters are predefined as the following.
The dc and ac spin [orbital] Hall conductivities ($\sigma_\mathrm{s[o]}(0)$ and
$\left. \partial \sigma_\mathrm{s[o]} (\omega) / \partial E \right|_{E_\mathrm{F}}$) are determined using the Kubo formula. 
We use a tight binding model to calculate the electronic state of Bi$_{1-x}$Sb$_x$\cite{liu1995prb,sahin2015prl,qu2023prb}.
The Kubo formula reads
\begin{equation}
\begin{aligned}
\sigma_{ji}^{k}(\omega)  &= i \hbar^2 e \sum_{\bm{k},n,m} \frac{f(\epsilon_n) - f(\epsilon_m)}{\epsilon_n - \epsilon_m}\frac{\langle u_n | \hat{v}_{j}^{k} | u_m \rangle \langle u_m | \hat{v}_{i} | u_n \rangle}{\varepsilon_n - \varepsilon_m + \hbar \omega + i \eta}.
    \label{eq:kubo}
\end{aligned}
\end{equation}
where $\hat{v}_i$ is the $i$th component of the velocity operator, $\hat{v}_j^k$ is the $j$th component of the spin (or orbital) current operator with spin (or orbital) direction along $k$. 
$\ket{u_n}$ is the Bloch eigenstate of an electron in band $n$ (with wave vector $\bm{k}$) and $\varepsilon_n$ is the corresponding eigenvalue.
$f(\epsilon)$ is the Fermi-Dirac distribution function and $\eta$ is a relaxation constant.
Here we use $\eta = 25$ meV and the temperature inside the Fermi distribution function is room temperature.
We employed a $k$-mesh with dimensions of $11 \times 11 \times 11$ to sample the irreducible Brillouin zone. 
In the main text, we use $\sigma_\mathrm{s[o]} (\omega) = \sigma_{zx}^y (\omega)$, where the spin [orbital] magnetic moment points along $y$.
$\sigma_{ij}^k (\omega)$ reduces to $\sigma_{ij}^k (0)$ in the limit of $\omega \rightarrow 0$.

The $x$-dependence of $\sigma_\mathrm{s[o]}(0)$ and $\left. \partial \sigma_\mathrm{s[o]}(\omega) / \partial E \right|_{E_\mathrm{F}}$ are shown in Fig.~\ref{fig:modelcal}(d-f). 
As is evident, $\sigma_\mathrm{s}(0)$ is nearly constant in the range of $x$ shown here, consistent with previous calculations\cite{sahin2015prl}.
$\sigma_\mathrm{o}(0)$, on the other hand, tends to decrease with increasing $x$.
The magnitude of $\sigma_\mathrm{o}(0)$ is nearly one third of $\sigma_\mathrm{s}(0)$ when $x=0$, in agreement with a previous report on Bi\cite{qu2023prb}.
Variation of $\left. \partial \sigma_\mathrm{s[o]}(\omega) / \partial E \right|_{E_\mathrm{F}}$ with $x$ is somewhat larger for the orbital component than that of the spin counterpart.
Note that the magnitude of $\left. \partial \sigma_\mathrm{s[o]}(\omega) / \partial E \right|_{E_\mathrm{F}}$ of Bi$_{1-x}$Sb$_{x}$ is more than one order of magnitude smaller than that of the transition metals\cite{marui2023prb}.

$\varepsilon_{xx}$ of Bi$_{1-x}$Sb$_{x}$ is determined by the refractive index $n$ and the extinction coefficient $k$, obtained using standard ellipsometry measurements.
The $x$ dependence of $n$ and $k$ are shown in Fig.~\ref{fig:transport}(d).
$\varepsilon_{xx}$ of the SiO$_@$ substrate is set to $n = 1.5, k = 0$.
We use $n = 1$, $k = 0$ for air and neglect the capping layer (1 nm thick oxidized Ti) for simplicity. 
$\rho_{xx}$ is obtained from Table~\ref{table:resistivity}.
$l_\mathrm{s}$ for Bi ($\sim18$ nm) was determined previously using transport measurements\cite{hirose2021prb}.
Here we fix $l_\mathrm{s}$ and $l_\mathrm{o}$ to 18 nm for all $x$.

The calculated MOKE signal is shown in Fig.~\ref{fig:modelcal}(a-c).
The right and left pointing triangles in Fig.~\ref{fig:modelcal}(a,b) show the $x$ dependence of the calculated MOKE signals ($\theta_\mathrm{K} / j$, $\eta_\mathrm{K} / j$) when contribution from the spin and orbital Hall effects are considered, respectively.
The squares represent the sum of the two contributions.
For $x=0$, the spin and orbital Hall effects tend to cancel out each other due to the difference in sign of $\sigma_\mathrm{SH}(0) \frac{\partial \sigma_\mathrm{SH}(\omega)}{\partial E}$ and $\sigma_\mathrm{OH}(0) \frac{\partial \sigma_\mathrm{OH}(\omega)}{\partial E}$, similar to the case in Pt\cite{marui2023prb}.
In contrast, the two effects add up for $x>0$.
The magnitude of the MOKE signal, $|\theta_\mathrm{K} + i \eta_\mathrm{K}| / j$, is plotted against $x$ in Fig.~\ref{fig:modelcal}(c).
The colored error bars attached to the squares show the range of $\theta_\mathrm{K}$ and $\eta_\mathrm{K}$ when $l_o$ is varied: the yellow and purple horizontal bars indicate the corresponding limits when $l_o = 1$ nm and $l_o = 100$ nm, respectively.
The calculated results roughly agrees with the experiments for $x=0$ and $x\sim0.28$ but are smaller when $x \sim 0.12$ and 0.2.
It is possible that contribution from the topological surface states, which is neglected in the model calculations, may influence the experimental results. 


\bibliography{ref_033026}
\end{document}